\documentclass[
preprint,
 amsmath,amssymb,
 aps,
showkeys,
showpacs
]{revtex4-2}

\usepackage{comment}
\usepackage{graphicx}
\usepackage{dcolumn}
\usepackage{bm}
\usepackage{hyperref}

\begin{document}

\preprint{MUPB/Conference section: }

\title{Review on reactor neutrino present and future}

\author{Thiago Bezerra}
 \altaffiliation{University of Sussex}
 \email{s26@sussex.ac.uk}
\affiliation{%
 Department of Physics and Astronomy, University of Sussex\\
Sussex House, Falmer, Brighton, BN1 9RH \\
United Kingdom 
}%


\date{\today}

\begin{abstract}
Nuclear reactors are an essential source of neutrinos. In this proceeding, I review the past and current status of the research on neutrino oscillations using reactor neutrinos. I also present a promising idea for positron tagging that will potentially be a game-changer in the field.
\end{abstract}

\maketitle

\section{Introduction}
\label{intro}
For my talk, a review of the research field on reactor neutrino detection for the 2021 edition of the Losomonov conference, I divided the contents into three parts: 1) A quick overview from the past; 2) The current status on reactor neutrinos, focusing on the standard three flavour neutrino oscillation and the inverse beta decay (IBD) reaction; and 3) What the future might bring us, showing a new exciting technology for positron tagging that has the potential to be a game-changer in the field.
There is rich physics with reactor neutrinos beyond oscillations and the IBD reaction.  Other speakers covered these topics, and the reader should check their talk/proceeding for further details (e.g. S.H. Seo on sterile neutrinos, K. Scholberg on coherent elastic neutrino-nucleus scattering, CE$\nu$NS, A. Parada on elastic scattering, and C. Guo on JUNO).

\section{Past}
\label{sec:past}
The detection of reactor antineutrinos marks the start of neutrino observation, when in the 50's we have the neutrino discovery by Reynes and Cowan using nuclear reactors as neutrinos source~\cite{Cowan1956}. Fast forward to late 70's, we have the first proposal on using neutrino detection to monitor the nuclear reactors~\cite{soviet}. In the 90's we started to use reactor neutrinos for oscillation searches. From that time there are two experimental highlights: 1) Bugey4, which had the best flux measurement for more than 20 years~\cite{bugey4}; and the CHOOZ experiment, which set the best limit on the neutrino mixing angle $\theta_{13}$~\cite{chooz} prior to its measurement in 2011. Finally, in the year 2000's we have KamLAND contributing to settle neutrino oscillations with its beautiful measurement of the neutrino reactor spectrum shape distortion driven by the solar parameters ($\theta_{12}$ and $\Delta m^{2}_{21}$)~\cite{kamland}.


\section{Reactor neutrinos and oscillation status}
\label{sec:present}

One of the reasons nuclear reactors are so popular in neutrino physics is that the reactors are a ``free'' and copious neutrino source. For example, via the fission chain reaction, a commercial fission reactor produces about $10^{20}$ neutrinos every second for each gigawatt of thermal energy ($GW_{th}$). The IBD reaction ($\bar{\nu}_{e} + p \rightarrow n + e^{-} $) is the main process we use to detect the reactor neutrinos and measure the neutrino oscillation parameter. The positron carries almost all of the neutrino energy and, tagging the neutron helps to discriminate against backgrounds.
Convoluting the reaction cross-section with the reactor neutrino spectrum, we have the characteristic shape of the detected spectrum (\textbf{figure}?), which integral is called mean cross-section per fission (MCSpF).

The CHOOZ results, cited in the previous section, triggered a worldwide race to measure the $\theta_{13}$ parameter~\cite{NPB2016}. If it was bigger than zero, or to set even more stringent limits, since the value of $\theta_{13}$ has crucial implications on neutrino oscillations (for example, if it turned out to be zero, current experiments searching for CP violations with neutrinos such as NOvA, T2K and DUNE, would not be viable). Although different groups proposed various experimental sites, all groups merged in the current three experiments~\cite{DB_18, RENO_18, DC_20}. Their main update concerning CHOOZ to measure $\theta_{13}$ is to place an additional detector near to the reactors, where almost no oscillations are present, while the usual detector is at the expected oscillation maximum. By comparing both measurements, we can cancel flux prediction and detection systematics that were the dominant ones for the CHOOZ experiment.

\begin{figure}[!tbh]
\includegraphics[width=0.9\textwidth]{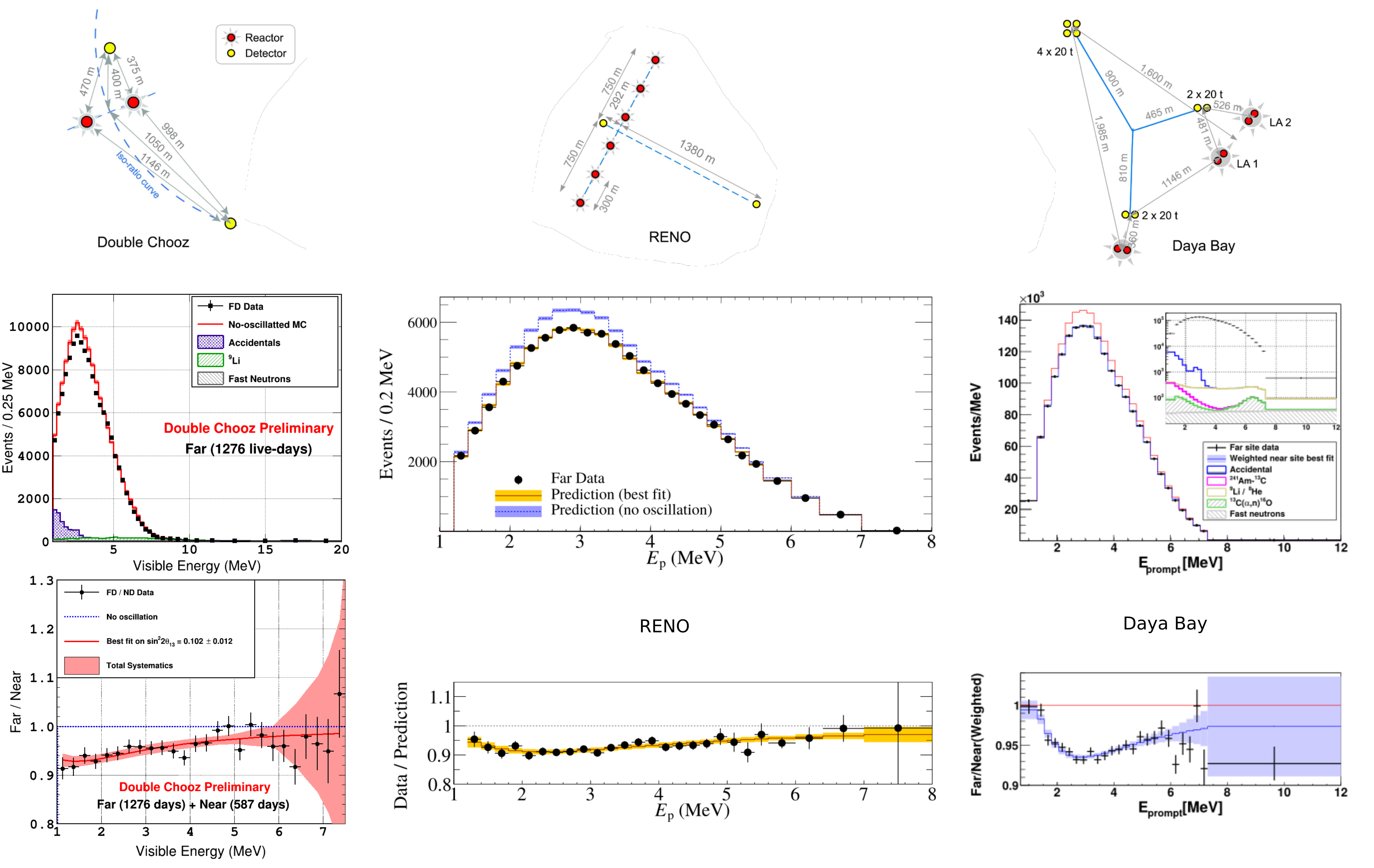}
\caption{\label{fig:spec}The experimental site, spectra and ratio of reactor $\theta_{13}$ experiments.}
\end{figure}

The top of Fig.~\ref{fig:spec} shows the reactor and detector arrangement of each experiment~\cite{reactorSite}, displaying their complementary set-ups of the number of reactors, detectors and different distances, meaning fractional flux contributions. The doctors are layered liquid scintillator volumes observed by photomultipliers, and the innermost volume is doped with gadolinium to maximize the IBD neutron detection and reduce accidental coincidences.

The data from the detectors provides a good energy reconstruction of the interactions. The middle of fig.~\ref{fig:spec} shows the measured spectra of each experiment's far detector, where we already see the missing event when compared to the expectation without oscillations. At the bottom part of the same figure is the far to near spectra ratio, showing the distortion due to the neutrino oscillations. This side by side comparison also shows the baseline effect, going from 1~km for Double Chooz, to about 1.4~km for RENO and 1.6~km for Daya Bay, shifting the position of the oscillation maximum, as expected.

All the current measurements of $\theta_{13}$, using different techniques, such as total neutron captures or neutron captures only on Gadolinium or Hydrogen, are in good agreement. They are also compatible with T2K measurement. There is also an agreement for the atmospheric mass splitting measured with reactors ($\Delta m^{2}_{ee}$) or accelerator ($\Delta m^{2}_{\mu\mu}$) neutrinos, where $\Delta m^{2}_{xx}$ is a linear combination of $\Delta m^{2}_{31}$ and $\Delta m^{2}_{32}$~\cite{nunokawa05}.

A by-product of the reactor $\theta_{13}$ experiments is the effort to understand the non-oscillated neutrino flux. This measurement brought a surprise that we have a deficit of about 6\%, now measured with a 1\% precision by Double Chooz~\cite{DC_20}, surpassing the Bugey4 result. Observing this deficit opened an opportunity that many experiments are exploring, the so-called very short baseline experiments. For details on the current status, see S.H. Seo talk/proceedings.

Additionally to the normalization that is not well-understood, the non-oscillated shape of the measured spectra also shows discrepancies when compared with the predictions, as you can see in \textbf{figure} X. The \textbf{figure} shows a similar distortion measured by different experiments. What is going on with the shape and normalization? The answer might be more strongly towards a problem with the prediction, or it is new physics. However, recent new re-evaluations of the prediction of the reactor neutrino flux have a better agreement with the data~\cite{giunti21}.

\section{Future: Positron tagging}
\label{sec:present}

The last part of my presentation talked about what the future with reactor neutrinos and IBD might bring to us (beyond the exciting JUNO project, explained in C. Guo talk/proceeding). An interesting scenario is the positron tagging.

Why positron tagging? First, we must remember the tremendous effort we have to go through for a clean IBD detection with reactor neutrinos. For a signal to the background of about ten to twenty, we need an underground detector with several layers of veto, shields and buffer volumes only to avoid things that are not positrons (the main background components, random coincidences, $^{9}Li/^{8}He$ decays, and fast-n, are all due to electrons, gammas, alphas or proton recoils). Moreover, remember that this detection technique is not so different from how reactor neutrino detection was done sixty years ago. Is it not a time for a change?

There are efforts to tag positrons with the current detectors. We have two examples from Double Chooz, exploring the powerful FADC's used to read the signal from all the PMTs. With the recorded waveform, we can see the fast and slow time components of the liquid scintillator light emission, for example. The first example is the ortho-positronium formation and decay observation~\cite{DC_posit}, which produces two peaks in the signal measured in the readout window (energy deposition with annihilation a few ns after). The downside of this method is the low efficiency achieved ($<50\%$). The second example is a pulse shape discrimination using a Fourier power spectra of the scintillator pulses, giving that its response is slightly different for electrons and positrons~\cite{DC_fourier}. The particle identification it produces is not satisfactory.

\begin{figure}[!tbh]
\includegraphics[width=0.8\textwidth]{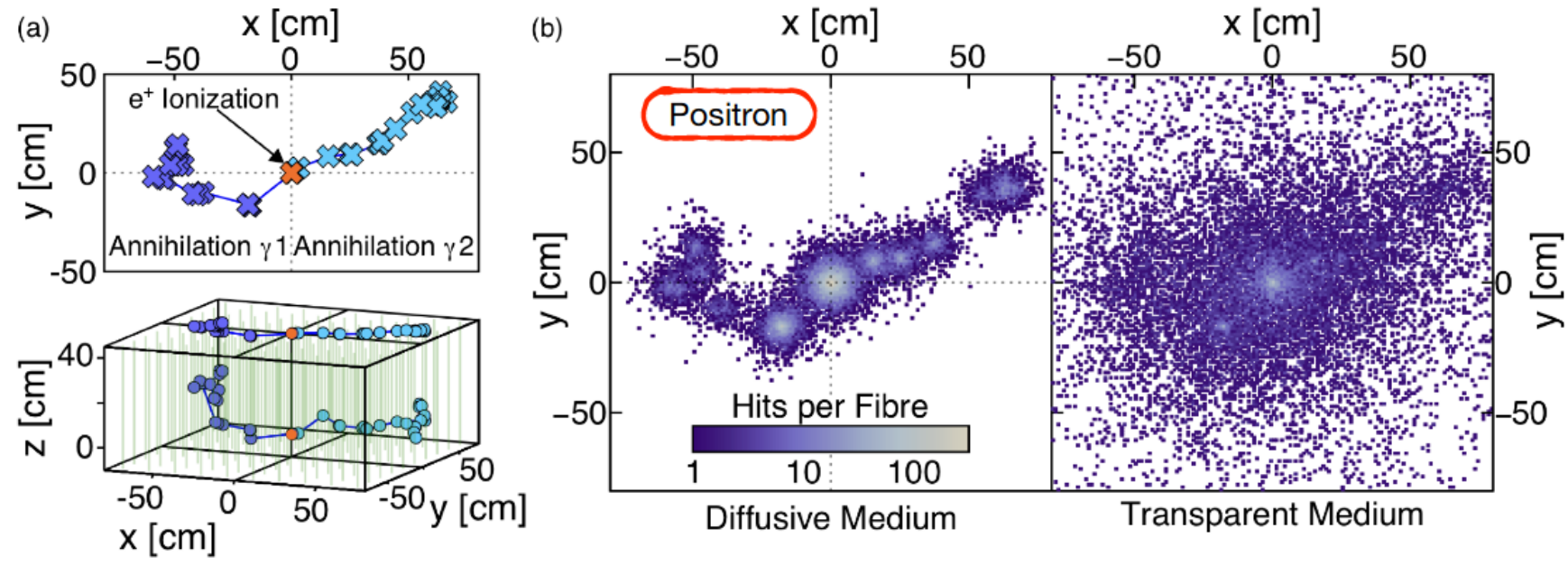}
\caption{\label{fig:opaque}Light hits in optical fibers for a transparent or opaque scintillator~\cite{liquido}.}
\end{figure}

A way to go beyond is with opaque scintillators. Consider a box detector with optical fibres running through it. Simulating a 2~MeV positron, we have the positron ionisation with the two annihilation gammas. Using the usual transparent liquid scintillator, we lost all the topological information of the positron interactions (energy deposition, annihilation and Compton scattering of the gammas), similar to what happens with a typical detector using PMTs. However, if we use an opaque scintillator, where opacity is through low scattering length and high absorption length of the scintillation light, we stochastically confine light near its creation point, preserving the precious topological information of particle interactions, as Fig.~\ref{fig:opaque} shows. We can distinguish now MeV gammas, electrons and positrons individually, which is unprecedented and astonishing.

A demonstration of a first-principle validation in the laboratory already exists~\cite{liquido,opaqueLS}. The possible impact on the reactor neutrino detection through the IBD reaction is truly amazing. By increasing the signal to the background to about one thousand, we are now talking about effective background free detectors. By killing the backgrounds just with the topology of the events, we do not need all the layers of defence (vetoes, shields, buffers) to avoid those things that are not positrons contaminating our sample. Therefore, we can increase the target mass at the current detector facilities by a factor of 10 at least.

\textbf{Impacts on $\theta_{13}$:} One of the impacts we have with this technology is the improvement of $\theta_{13}$ precision. Currently, we do not expect it to get better than 2\% uncertainty after all the data of the current experiments are released. Other projects such as DUNE will not improve it significantly either. Now we can get a sub-par cent uncertainty using available technologies (optical fibres, like NOvA, with silicon photomultipliers). Preliminary results show that the main systematic of concern will be the energy scale, as is for JUNO~\cite{superChooz}. This precise measurement will be a key for the discrimination of models and understanding the funny nature of $\theta_{13}$, much smaller than the other two mixing angles, but also to add further constraints in the Jarlorg invariant, together with the measurements of JUNO, DUNE, Hyper-K and so on. For example, with $\theta_{13}$ measured with sub-\% precision, DUNE can improve its sensitivity on CP violation equivalent to a few years of data taking.
This improvement is a natural extension of the current synergy that already happens with T2K and NOvA, benefiting from a precise $\theta_{13}$ measurement of the reactor neutrino experiments.

\section{Summary}

In this talk, I tried to show that nuclear reactors are a pivotal source for neutrino physics. The inverse beta decay channel drives the $\theta_{13}$ measurement, which is the oscillation parameter measured with the lowest uncertainty to date. $\theta_{13}$ uncertainty will not be the best after JUNO, which will measure $\theta_{12}$, $\Delta m^{2}_{21}$ and $\Delta m^{2}_{31}$  with sub-\% precision.
Positron tagging with an opaque scintillator is a counter-intuitive detection technique, having a promising future. A proof-of-principle exists, and there are works ongoing towards a multi-ton demonstrator. Such detector opens up a possible window for a sub-\% measurement of $\theta_{13}$, which will inevitably have consequential impacts, such as a better measurement of the CP violation in the neutrino sector.




\end{document}